\documentclass[runningheads]{llncs}

\usepackage{graphicx}

\begin{document}

\title{Identification of spatial dynamic patterns of behavior using weighted Voronoi diagrams}

\titlerunning{Identification of spatial dynamic patterns of behavior using WVD}

\author{
Martha Lorena Avendaño-Garrido\inst{1}\orcidID{0000-0001-7956-8958} 
\and Carlos Alberto Hernández-Linares \inst{1}\orcidID{0000-0003-3786-0373}
\and Brenda Zarahí Medina-Pérez\inst{1}\orcidID{0009-0002-0391-7722} 
\and Varsovia Hernández \inst{2}\orcidID{0000-0002-6849-8461} 
\and Porfirio Toledo \inst{1}\orcidID{0000-0001-9639-446X} 
\and Alejandro León \inst{2}\orcidID{0000-0001-7386-9784}
}

\authorrunning{M. L. Avendaño-Garrido et al.}

\institute{
Facultad de Matemáticas, Universidad Veracruzana, Xalapa, Ver., México 
\and Centro de Investigaciones Biomédicas, Universidad Veracruzana,  Xalapa, Ver., México \\
\email{carlhernandez@uv.mx}}

\maketitle

\begin{abstract}
This study proposes an innovative approach to analyze spatial patterns of behavior by integrating information in weighted Voronoi diagrams. The objective of the research is to analyze the temporal distribution of an experimental subject in different regions of a given space, with the aim of identifying significant areas of interest. The methodology employed involves dividing the experimental space, determining representative points, and assigning weights based on the cumulative time the subject spends in each region. This process results in a set of generator points along with their respective weights, thus defining the Voronoi diagram. The study also presents a detailed and advanced perspective for understanding spatial behavioral patterns in experimental contexts.

\end{abstract}

\section{Introduction}
The Spatial Dynamics Behavior Analysis (SDBA) focuses on analyzing the displacement of individuals, whether animals or humans, within the context of various behavioral phenomena such as learning, motivation, and fear, among others (see \cite{leon2021}). One of the inherent challenges of SDBA involves visualizing the quantitative relationships between spatial variables associated with the phenomena of interest and other relevant variables. In this context, SDBA requires an analysis and visual representation method capable of segmenting space into zones defined by organism behavior, thus enabling the analysis of Regions of Behavioral Relevance (RBR) and the representation of the emergence and evolution of RBR over different time points. In this study, we propose employing weighted Voronoi diagrams to address this challenge.

A Voronoi diagram is constituted by regions that divide the space according to a finite set of points called generators. Each region is associated with a single generator point and each point in the space is associated with at least one region (see \cite{boots2009,gallier2017,preparata2012}). When, in addition to considering the location of the generator points, a weight is incorporated, we call them weighted Voronoi diagrams. In this paper, we focus specifically on the multiplicative weighting. It is important to mention that between the 1950s and 1970s, multiplicatively weighted Voronoi diagrams were widely used to solve problems in market and urban analysis (see \cite{mu2004}). 

Voronoi diagrams have been highlighted as a versatile tool for exploring diverse aspects of behavioral spatial dynamics. They have been used in various fields such as urban mobility modeling to determine population concentration, travel speed, and direction (see \cite{Manca2017,Viloria2020}). Similarly, in sports science, they have been useful in characterizing collective behavior, identifying crucial moments in games, and analyzing spatial organization (see \cite{Eliakim2022,Gudmundsson2018}). However, in these contexts, the potential of weighted Voronoi diagrams has not been fully exploited, limiting their use to movement classification or the determination of relevant regions in a single episode or session. In contrast, in this work, we use weighted Voronoi diagrams to analyze the change in dynamics over multiple sessions in order to describe the evolution of the individual behavior under a given spatiotemporal dynamics of the environment.

\section{Weighted Voronoi diagrams}
We will consider a generator set $P = \lbrace p_1,p_2,\ldots,p_n \rbrace \in \bbbr^2$, where each point $p_i$ is assigned a weight $w_i>0$, represented by the parameter set
$ W=\lbrace w_1,w_2,\ldots,w_n \rbrace.$
It is considered that the weight $w_i$ reflects the ability of the generator $p_i$ to influence the space.
This weight is used to define a weighted distance relative to $p_i$, denoted as $d_w(x, p_i)$. It is important to note that we will call this a distance, even though it is not in the strict sense of the definition of a metric space. In this context, we define the following.

\begin{definition}
The weighted Voronoi polygon associated to a point $p_i\in P$ is the region defined by
$$Vor_w(p_i) = \lbrace x \in \bbbr^2 :  d_w(x,p_i)\leq d_w(x,p_j), \forall p_j \in P \rbrace. $$
\end{definition}

\begin{definition}
A weighted Voronoi diagram of $P$ is a partition of $\bbbr^2$ into $n$ regions, defined by the Voronoi polygons associated to set $P$, that is 
$$\{Vor_w(p_1), \ldots, Vor_w(p_n)\}.$$
The Voronoi diagram of $P$ is usually denoted as 
$$Vor_w(P) = \bigcup_{i=1}^n Vor_w(p_i).$$ 
\end{definition}

\begin{definition}
For $i \neq j$, the domain region of $p_i$ over $p_j$ is defined as $$ Dom(p_i,p_j) = \lbrace x \in \bbbr^2 : d_w(x,p_i) \leq d_w(x,p_j) \rbrace.$$
So, we have
$$Vor(p_i) = \bigcap_{j\neq i, j=1}^n Dom_w(p_i,p_j).$$
\end{definition}

\begin{definition}
The weighted bisector between $p_i$ and $p_j$, with $j\neq i$, is defined as
$$ b_w(p_i,p_j) = \lbrace x \in \bbbr^2 : d_w(x,p_i) = d_w(x,p_j) \rbrace.$$    
\end{definition}

Note that the domain regions of $p_i$ and $p_j$ correspond to spaces divided by their bisector, so we can express this set as
$$b_w(p_i,p_j)=Dom_w(p_i,p_j)\cap Dom_w(p_j,p_i).$$

\begin{definition}
Given $p_i$ and $p_j$, with $j\neq i$, its weighted Voronoi edge $e_w(p_i,p_j)$ is the intersection of its polygons if it is not empty and has more than a single point; that is, $$e_w(p_i,p_j) = Vor_w(p_i) \cap Vor_w(p_j).$$
We will call the union of weighted Voronoi edges a Voronoi lattice and the intersection of three or more Voronoi polygons a Voronoi vertex.
\end{definition}

In the literature, four variants of weighted Voronoi diagrams are distinguished: multiplicatively weighted, additively weighted, compound weighted, and power weighted. The choice among them depends on how the values of $W$ are used in the function $d_w$ (see \cite{boots2009}).

In this paper, we will use multiplicatively weighted Voronoi diagrams. In particular, we will consider $w_i>0$, for $i=1,\ldots,n$, and we will use the weighted distance defined by $$d_{w}(x,p_i)= \frac{\|x-p_i\|}{w_i}, $$ where $\|x - p_i\|$ represents the Euclidean distance between $x$ and $p_i$. Thus, the domain region of $p_i$ over $p_j$ is
\[Dom_{w}(p_i,p_j)   = \left \lbrace x \in \bbbr^2 : \frac{\|x-p_i\|}{w_i} \leq \frac{\|x-p_j\|}{w_j} \right\rbrace.\]
Note that when $w_i=w_j$, the domain corresponds to those points which are closer to $p_i$ than to $p_j$. Actually, when $w_i = w_j$ for all $i,j \in \{1, \dots, n\}$ the weighted Voronoi diagrams coincide with the classical Voronoi diagrams, and every Voronoi polygon is the set of points closer to $p_i$ than to another $p_j$, with the euclidean metric. On the other hand, if $w_i < w_j$, the domain region of $p_i$ over $p_j$ can be expressed as 
\[
Dom_{w}(p_i,p_j) = \left\lbrace x \in \bbbr^2  :  \| x - o \| \leq \epsilon \right\rbrace ,
\]
where 
\[
o = \frac{w_j^2}{w_j^2- w_i^2} p_i - \frac{w_i^2}{w_j^2- w_i^2} p_j, \qquad 
\epsilon = \frac{w_iw_j}{w_j^2-w_i^2}\|p_i-p_j\|.
\]
Then $Dom_{w}(p_i,p_j)$ is a closed ball with center at $o$ and radius $\epsilon$. Whereas for the case $w_i > w_j$, the domain region of $p_i$ over $p_j$ is  
$$ Dom_{w}(p_i,p_j) = \left\lbrace x \in \bbbr^2 : \| x - o \| \geq \epsilon \right\rbrace, $$
which defines the complement of the open ball with center at $o$ and radius $\epsilon$. 
Thus, if $w_i \neq w_j$, the bisector between $p_i$ and $p_j$ can be written as 
\[ b_w(p_i,p_j) = \left\lbrace x \in \bbbr^2 : \left|\left| x- o \right|\right| = \epsilon \right\rbrace.\]
That is, in this case, the bisector is a circle.  

In the literature we find the following result characterizing the geometry of the multiplicatively weighted Voronoi lattice (see \cite{boots2009}).

\begin{theorem}
The edges of a multiplicatively weighted Voronoi region are circular arcs if and only if the weights of two adjacent regions are not equal, and are straight lines in the plane if and only if the weights of two adjacent regions are equal. 
\end{theorem}

Due to the wide range of applications of Voronoi diagrams, several algorithms have been developed for their generation, including techniques such as Domain Intersection, the Incremental Algorithm, and the Divide and Conquer approach, and others (see \cite{aurenhammer2013,boots2009,deBerg2008}).

\begin{figure}
\centering
\includegraphics[scale=0.17]{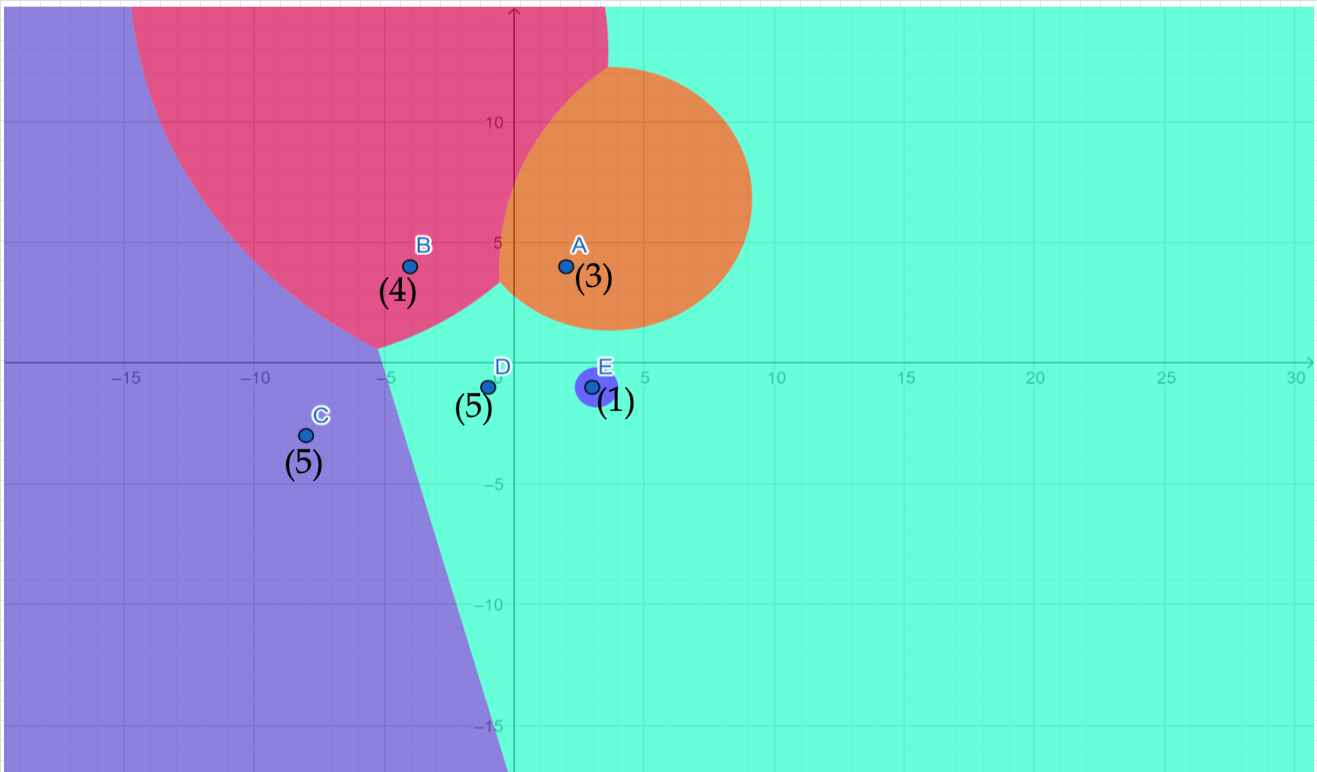}
\includegraphics[scale=0.174]{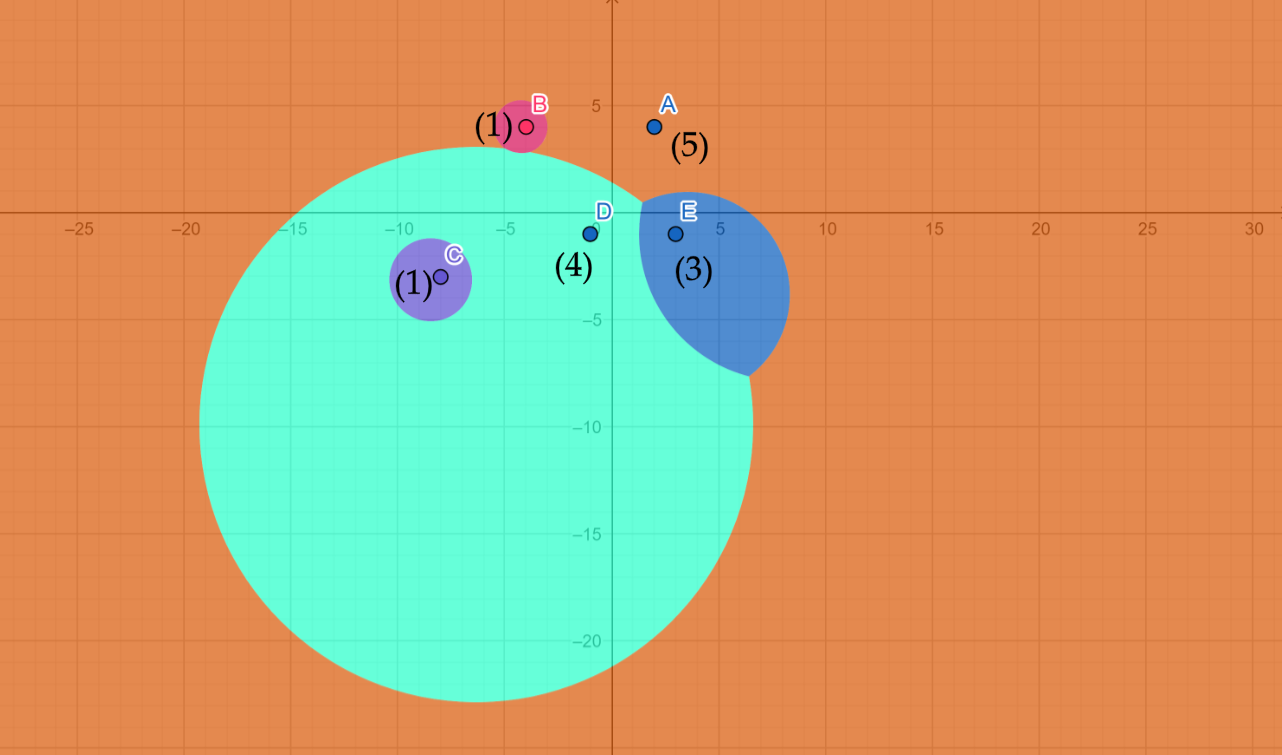}
\caption{Multiplicatively weighted Voronoi diagrams, the numbers represent the corresponding weights.}
\label{VoronoiDP}
\end{figure}

In Fig. \ref{VoronoiDP} we can see that multiplicatively weighted Voronoi regions are not necessarily convex.

\section{Application for SDBA}
Behavior Analysis constitutes a discipline that aims to understand the behavior of individuals, exploring both the behavior itself and the variables that influence it, including the spatiotemporal dynamics of the environment. Within this field, the study of organism movement, within the framework of behavioral phenomena and repertoires, addressed from perspectives such as organization, development, and change, is termed Spatial Dynamics Behavior Analysis (see \cite{leon2020ecological}).

In SDBA, one of the main challenges lies in visualizing quantitative relationships between spatiotemporal variables of the environment and spatiotemporal features of behavior. Specifically, a clear representation of the organization, development, and evolution of interactions between variables, such as the time of stay in a specific zone, and treatment variables, such as the type of reinforcement schedule defined by the spatial-temporal constancy or variation in water delivery, is sought. This approach is applied to the study of specific behavioral phenomena, such as spatial differentiation in water-seeking situations in water-deprived rats.

Next, we will explore the use of weighted Voronoi diagrams in this context, starting with a detailed description of the experimental process, the type of data collected, and the data collection process.

\subsection{Data Acquisition}
The study was conducted by members of the Comparative Psychology Laboratory at Universidad Veracruzana within the facilities of the W. N. Schoenfeld Laboratory. The experimental subjects were 12 Wistar rats subjected to 22 hours of water deprivation, with continuous access only to food.

A single-case experimental design was implemented within established behavior analysis protocols, but incorporating a high-resolution behavioral record for each rat (7200 data points per session). For data collection, a modular displacement chamber was employed, with an experimental space measuring $92\times 92\times 33$ cm in length, width, and height. Within this experimental space, four limited-availability water dispensers were placed, one in the middle of each wall of the chamber. The provided data indicate the subject's trajectory in coordinates $(x,y)$ over time, recorded at a resolution of $5$ frames per second.

Fifty sessions, each lasting 20 minutes, were conducted for these experiments. The studies involved the intersection of two environmental features (Fixed and Variable) within two environmental dimensions (Time and Space):
\begin{enumerate}
    \item[FT:] \textit{Fixed Time}, water delivery occurred every $30$ seconds, with an availability window of $3$ seconds. 
    \item[VT:] \textit{Variable Time}, water delivery time varied, within an average of $30$ seconds, with an availability window of $3$ seconds.
    \item[FS:] \textit{Fixed Space}, water was delivered consistently at the same dispenser in all sessions, located at the coordinate $(50,0)$.
    \item[VS:] \textit{Variable Space}, water was randomly delivered to one of the four dispensers.
    \label{con:tiempos}
\end{enumerate}

Thus, four experiments were obtained: Fixed Time - Fixed Space (FT-FS); Fixed Time - Variable Space (FT-VS); Variable Time - Fixed Space (VT-FS); and Variable Time - Variable Space (VT-VS). Each experiment was conducted with three subjects for each combination. 

\begin{figure}[ht]
 \centering
 \begin{tabular}[b]{c}
\includegraphics[width=0.235\textwidth]{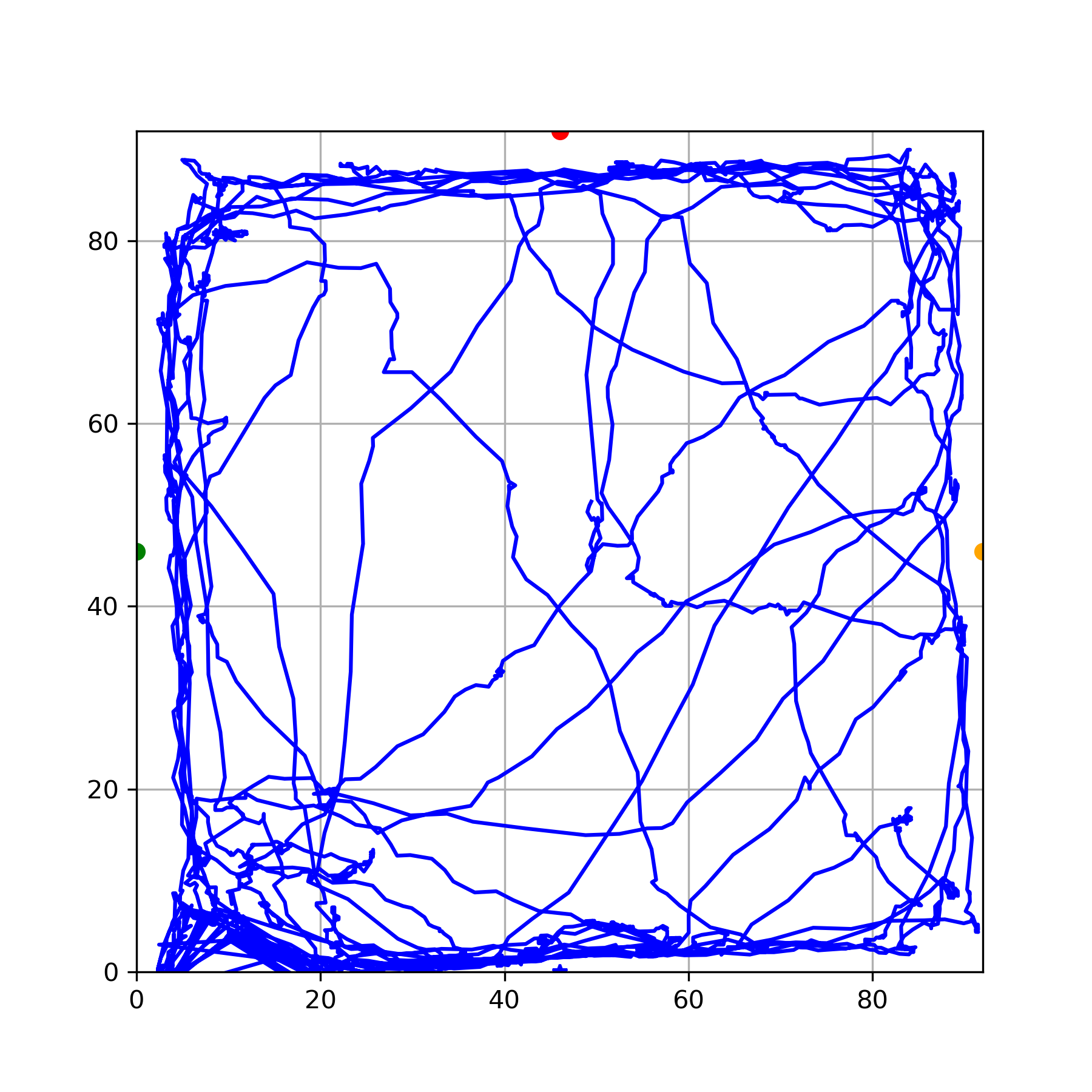}\\
\small (a)
\end{tabular}
\begin{tabular}[b]{c}
\includegraphics[width=0.235\textwidth]{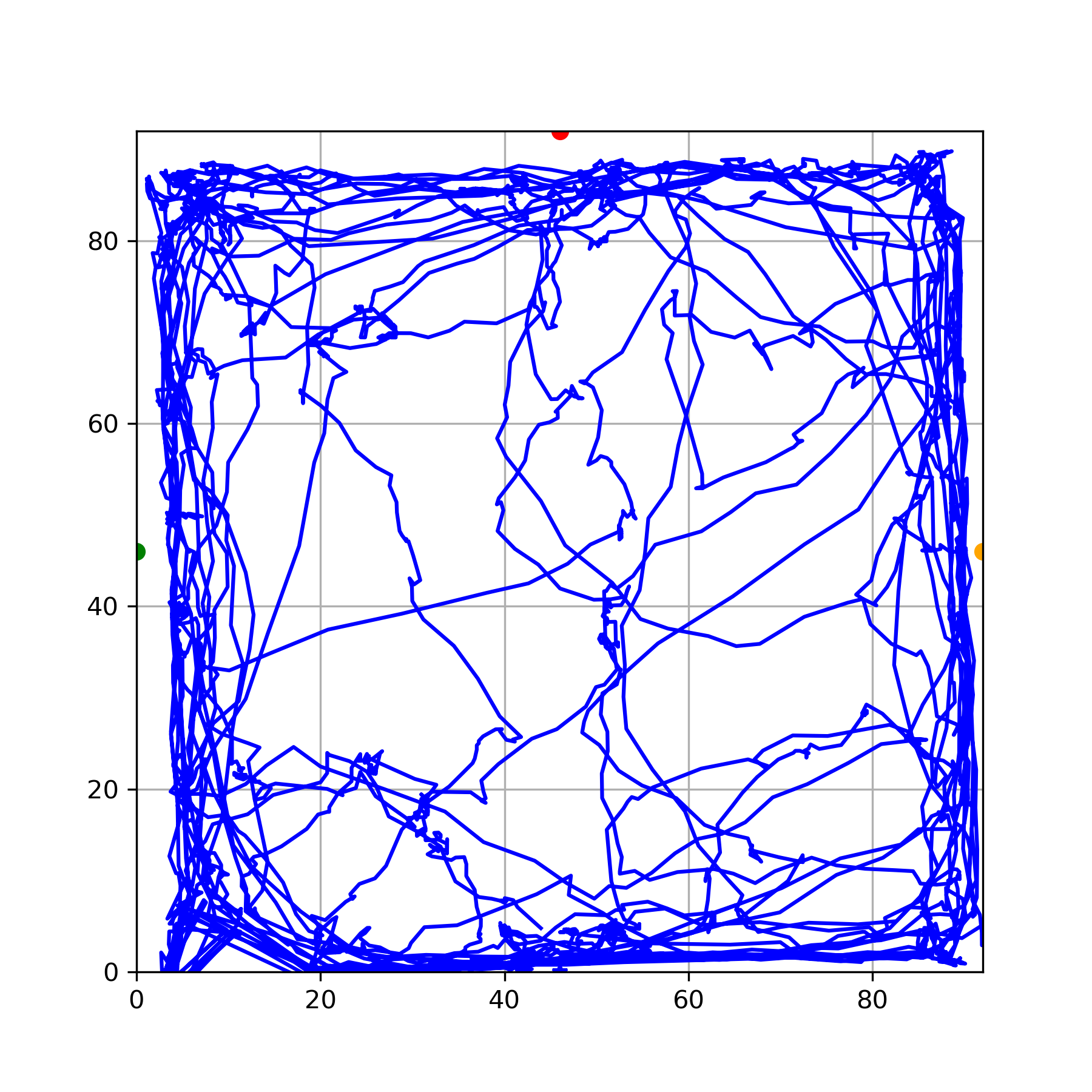}\\
\small (b)
\end{tabular} 
\begin{tabular}[b]{c}
\includegraphics[width=0.235\textwidth]{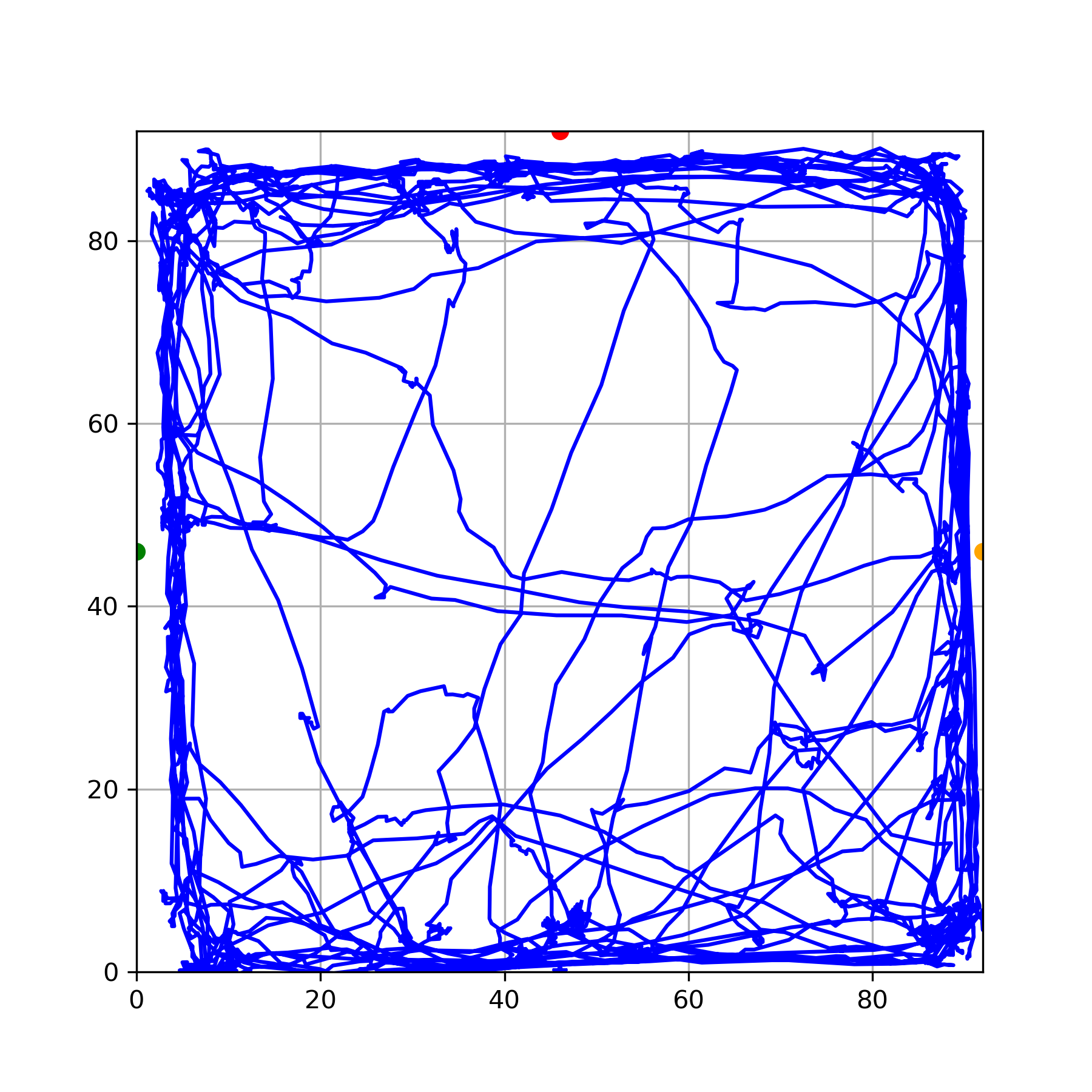}\\
\small (c)
\end{tabular}
\begin{tabular}[b]{c}
\includegraphics[width=0.235\textwidth]{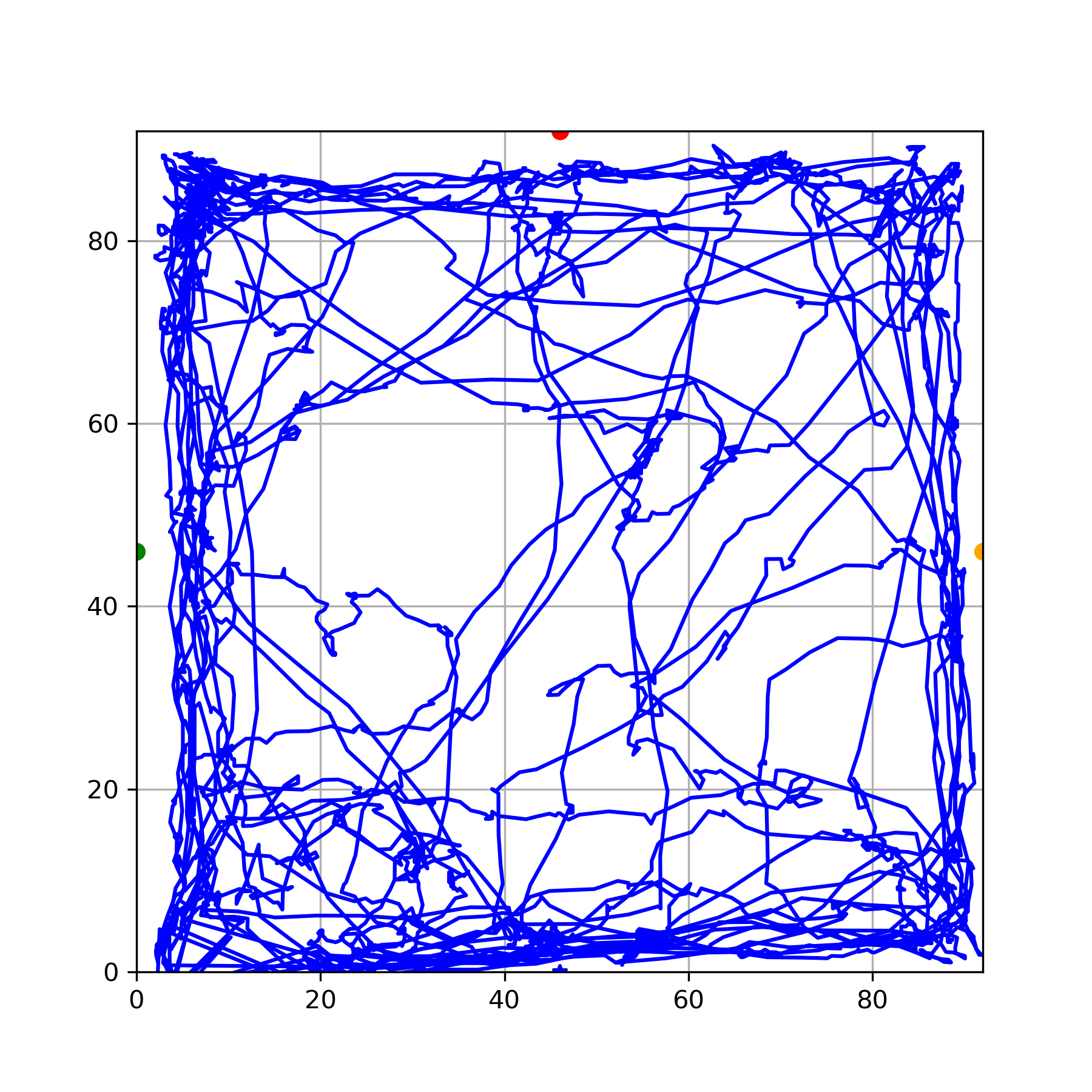}\\
\small (d)
\end{tabular}
\caption{Subjects' Trajectory.}
\label{fig:ruta}
\end{figure}

Figure $\ref{fig:ruta}$, illustrates an example of trajectories followed by some subjects in the first session for combinations FT-FS (Figure \ref{fig:ruta} (a)), FT-VS (Figure \ref{fig:ruta} (b)), VT-FS (Figure \ref{fig:ruta} (c)) and VT-VS (Figure \ref{fig:ruta} (d)).

\subsection{Generation of weighted Voronoi diagrams}

In order to be able to construct the weighted Voronoi diagrams with the experimental data collected, it is necessary to have a generator set up and to assign the appropriate weights to each of its points. The procedure to achieve this construction is detailed below.

First, the experimental space was divided into $n \times m$ regions of uniform size and the center of each of them was calculated, thus obtaining a set of $n \times m$ points, which we will call \textit{initial set}. Once this division was obtained, the software \textbf{MOTUS}\footnote{ MOTUS, is a software that allows to generation of different graphical representations of data from the displacement of an individual, recorded as changes in $(x,y)$ coordinates over time. } was used to obtain the \textit{cumulated time} by region, corresponding to the cumulative number of times the subject was in each area (see \cite{leon2020motus}). 
This information can be visualized as a matrix.

The following procedure was used to build the generator set and assign weights to each of its points:
\begin{enumerate}
    \item Each point in the initial set was assigned its weight based on its cumulative time.
    \item The generator set $P$ was defined as the set of points whose assigned weights were non-zero.
    \item The set $W$ contains all the weights corresponding to the generating set $P$.
\end{enumerate}
This procedure guarantees an accurate representation of the generator points and their respective assigned weights, providing the necessary elements for the generation of the weighted Voronoi diagrams. 
It can be observed that \textit{initial set} and \textit{cumulative time} in the region are fully determined, the first one at the moment when the parameters for the analysis are configured and the second one by data acquired from the behavioral experiment. This implies that there is no randomness in the diagram they generate.
To accomplish the above, the domain intersection algorithm was implemented and executed in Python.

\begin{figure}[ht]
    \centering
\includegraphics[width=0.3\textwidth]{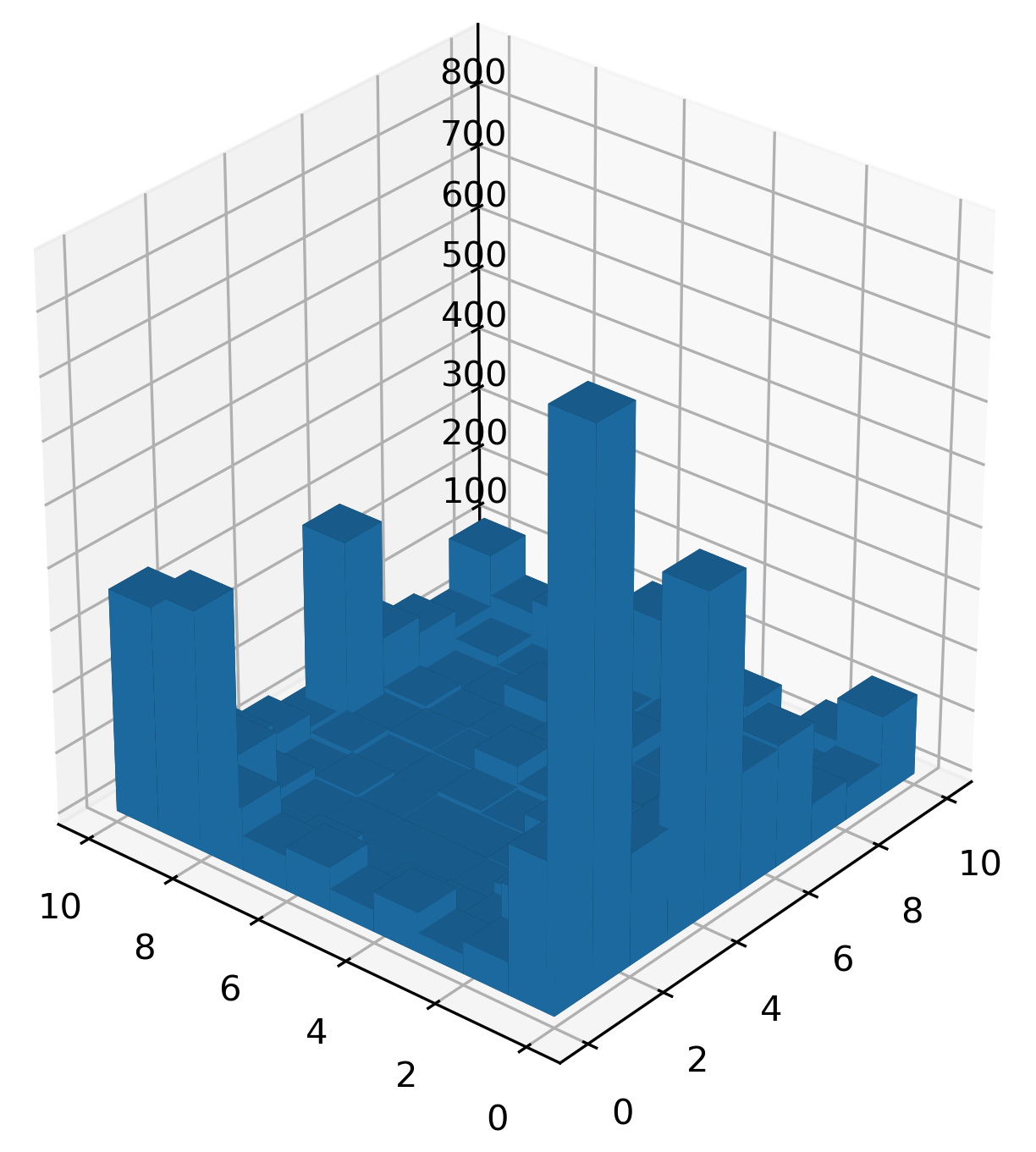}    \includegraphics[width=0.31\textwidth]{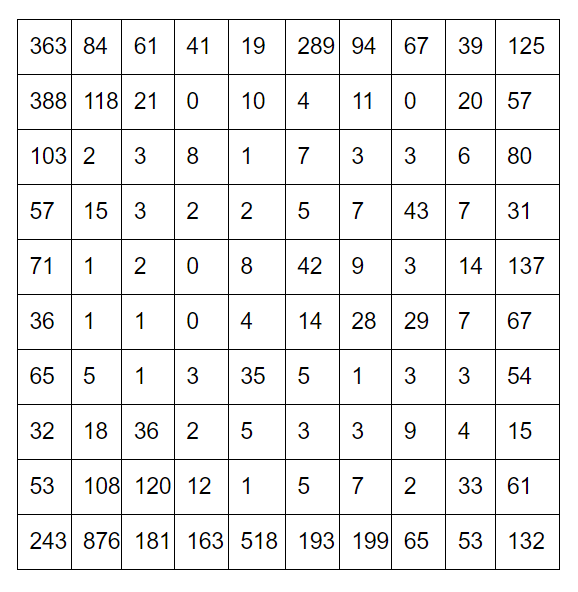} 
\includegraphics[width=0.32\textwidth]{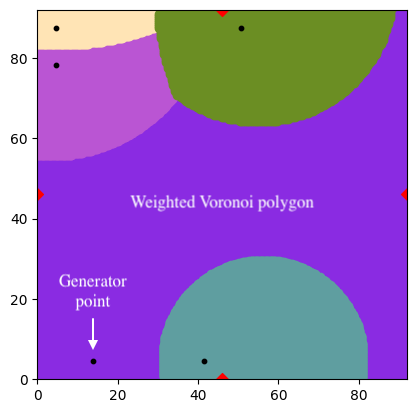}   
\caption{Accumulated time in regions and the corresponding multiplicatively weighted Voronoi diagram.}
\label{fig:Regi}
\end{figure}

Fig. \ref{fig:Regi} shows the data derived from the process described above, in which the experimental space, with dimensions of $92$ cm $\times 92$ cm, was partitioned into a grid of $10 \times 10$, generating an initial set of $100$ points, each with its associated weight.
In the multiplicatively weighted Voronoi diagram in Fig. \ref{fig:Regi}, the red triangles indicate the location of the water dispensers, and the color-determined region points to the weighted Voronoi region of the generator points, represented by a black dot. The regions shown are those corresponding to the generators with the highest weights, as they are arranged in descending order. In order to streamline the visualization of the identified regions in this study, only the first five regions with the highest weights were included.

\subsection{Data Analysis}
To illustrate the use of Voronoi diagrams in SDBA, four subjects were selected from the twelve experimental subjects, with each subject exposed to one of the four aforementioned conditions. Tables \ref{Tabla: 2-sesiones} and \ref{Tabla: 3-sesiones}, display the condition, subject, and five sessions in which the weighted Voronoi diagram of the five generators with the highest weight was generated. The Voronoi diagrams show the behavioral segmentation of the experimental space. The sessions were chosen to depict the evolution of behavioral spatial segmentation throughout the experiment—highlighting the contrast between the transition in early sessions (1 to 4) and a later session that illustrates a typical steady behavioral state. This detailed analysis of behavioral spatial evolution serves as an example of SDBA.

It is noticeable the similarity of the diagrams in session 1 between the four experiments. In all cases, the generators with more time spent were distributed between corners and dispenser zones, as Regions of Behavioral Relevance (RBR), with both circular arcs and straight-line arcs. Nevertheless, as experiments pro\-gress, different segmentations emerge between them as follows.

It is noteworthy that, for Subject 1 under the condition \textit{Fixed Time-Fixed Space}, the generators where the subject spent the most time from session two onward are near the dispenser. These RBRs are represented in grayish blue and neon green colors, and such RBRs are also present in the last session of the experiment. Starting from session two, all RBRs are circular arcs, and the straight-line arcs disappear completely, depicting a dominance of certain regions or inequality between regions.
\begin{center}
\begin{table}[ht]
    \begin{tabular}{|c|c|c|c|c|c|}\hline
 & Session 1 & Session 2 & Session 3 &Session 4 & Session 20 \\\hline
FT-FS&&&&&\\
Subject 1 & 
\includegraphics[scale = 0.19]{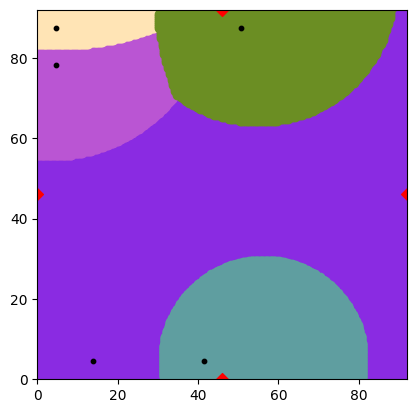} & 
\includegraphics[scale = 0.19]{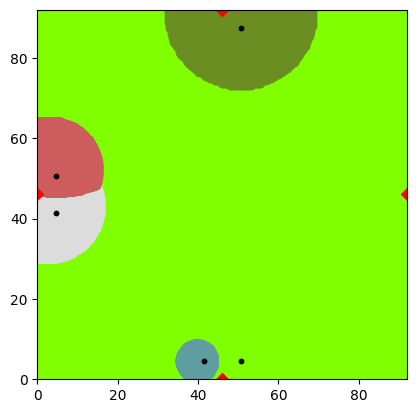} & 
\includegraphics[scale = 0.19]{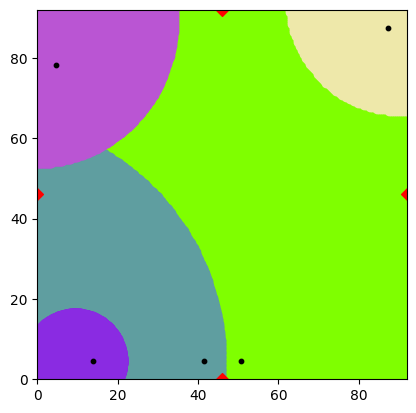} & 
\includegraphics[scale = 0.19]{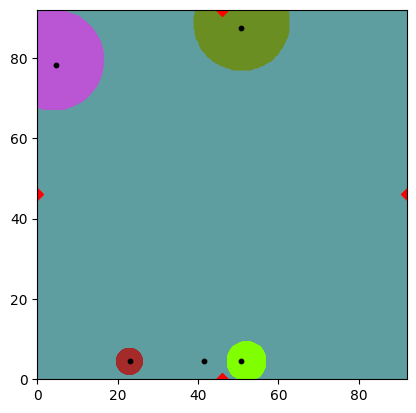} &
\includegraphics[scale = 0.19]{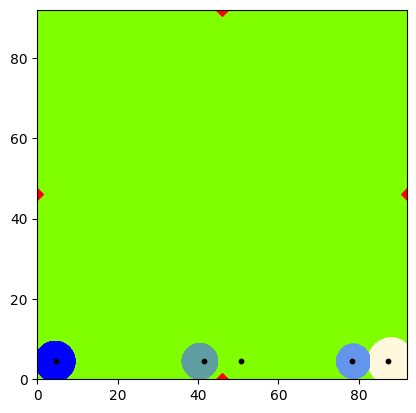} \\\hline
VT-FS&&&&&\\
Subject 7 & 
\includegraphics[scale = 0.19]{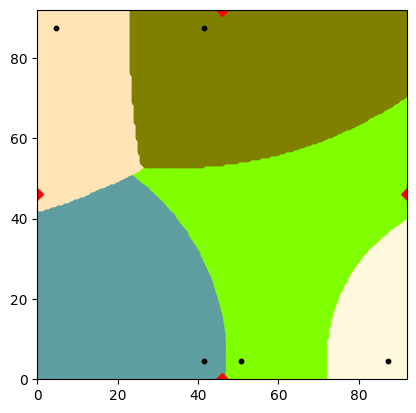} & 
\includegraphics[scale = 0.19]{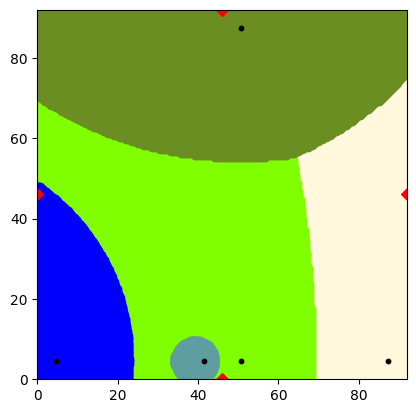} & 
\includegraphics[scale = 0.19]{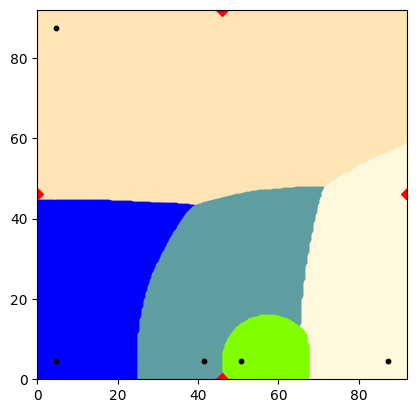} &
\includegraphics[scale = 0.19]{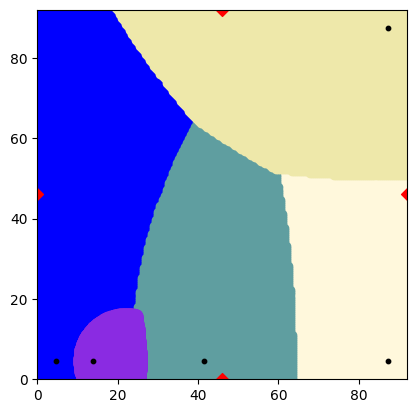} &
\includegraphics[scale = 0.19]{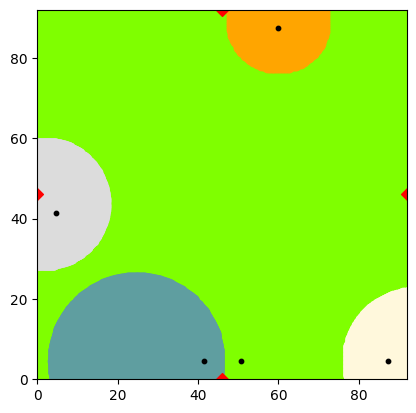} \\\hline
\end{tabular}
    \caption{Multiplicatively Weighted Voronoi diagrams of subjects 1 and 7.} \label{Tabla: 2-sesiones}
\end{table}
\end{center}
On the other hand, for Subject 7 under the condition \textit{Variable Time-Fixed Space}, although the water delivery is located in the same zone as in the condition of the first subject, the diagrams vary significantly. In this case, no single dominating region is identified across sessions, regardless of the water delivery only being in one region. The simple variability in the time delivery induced more varied RBRs, including corners and dispenser zones. Additionally, in comparison to Subject 1, both circular and some straight-line arcs can be observed in the first four diagrams related to the transition sessions, but only circular arcs for session twenty.

In contrast, Subject 4 under the condition \textit{ Fixed Time-Variable Space}, was expected to visit each of the dispensers. From session two onward, a notable progression toward symmetric regions was observed, with straight-line arcs becoming especially prominent in session 4.

Finally, in the sessions corresponding to Subject 10 under the condition with higher variability, \textit{Variable Time-Variable Space}, two differentiated kinds of diagrams are observed. Sessions 1-3 revealed an RBR distribution trending toward equity, characterized by prominent straight-line arcs, as expected under the Variable Space condition. However, a sudden shift occurred in session 4, where generators with more time spent were predominantly located near the bottom and left dispensers, as well as the bottom-right corner. In this scenario, RBRs were now delimited by circular arcs. Notably, in session 20, all weight generators were concentrated at the bottom of the experimental space. This transition from an equitable segmentation to a dominant segmentation, as indicated by the size of RBDs and the features of the arcs, may be attributed to heightened uncertainty in water delivery conditions resulting from the variability in both Space and Time.
\begin{center}
\begin{table}[ht]
    \begin{tabular}{|c|c|c|c|c|c|}\hline
 & Session 1 & Session 2 & Session 3 &Session 4 & Session 20 \\\hline
FT-VS&&&&& \\ 
Subject 4& 
\includegraphics[scale = 0.19]{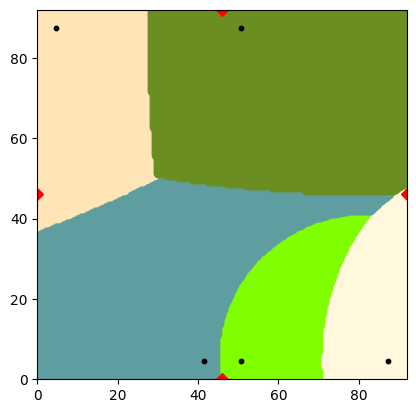} & 
\includegraphics[scale = 0.19]{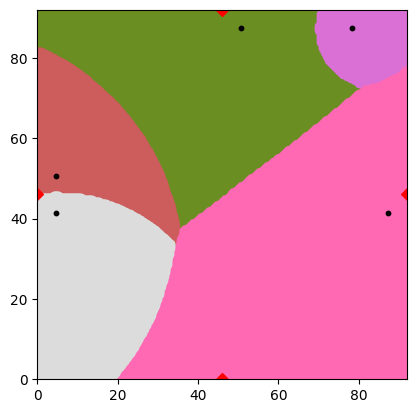} & 
\includegraphics[scale = 0.19]{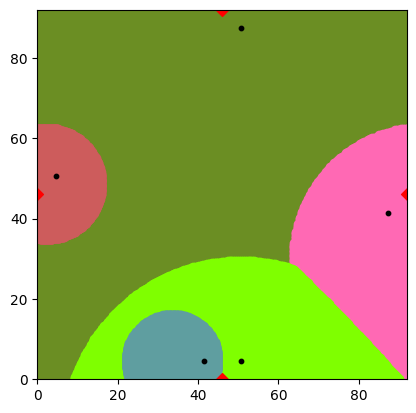} &
\includegraphics[scale = 0.19]{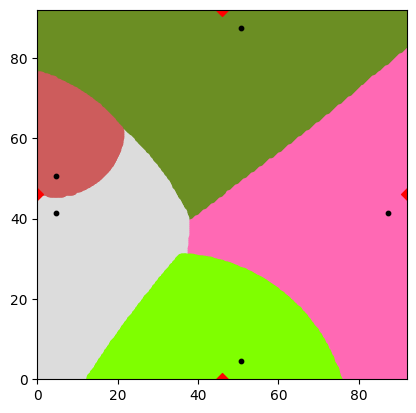} &
\includegraphics[scale = 0.19]{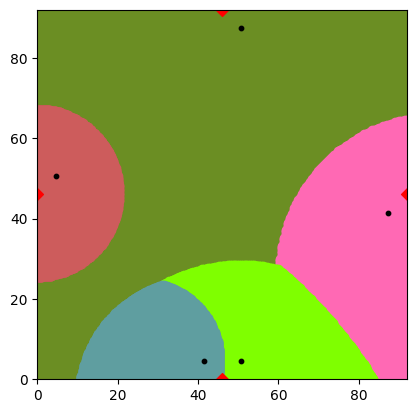}  \\\hline
VT-VS &&&&& 
\\Subject 10 & 
\includegraphics[scale = 0.19]{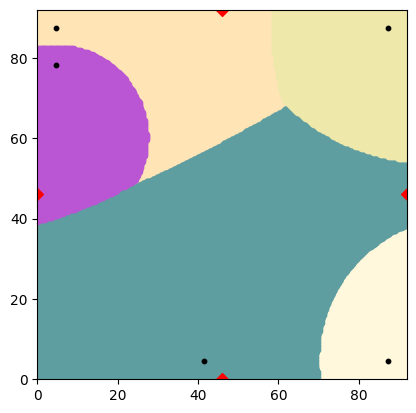} & 
\includegraphics[scale = 0.19]{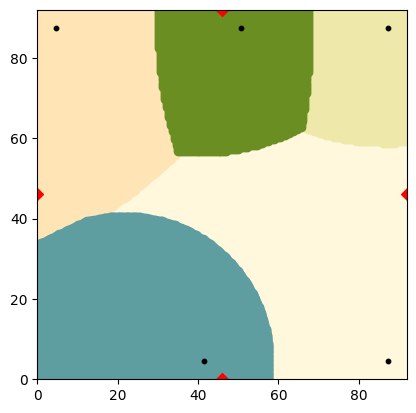} & 
\includegraphics[scale = 0.19]{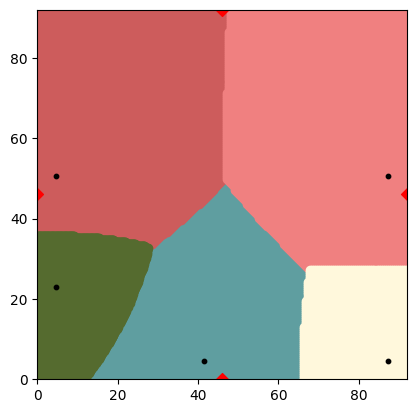} &
\includegraphics[scale = 0.19]{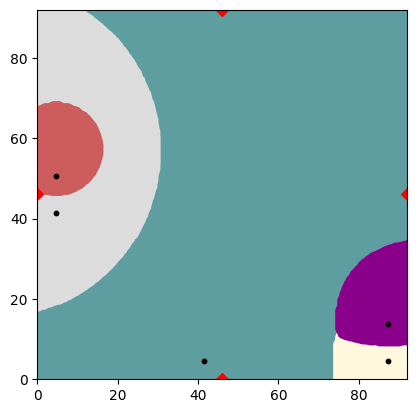} &
\includegraphics[scale = 0.19]{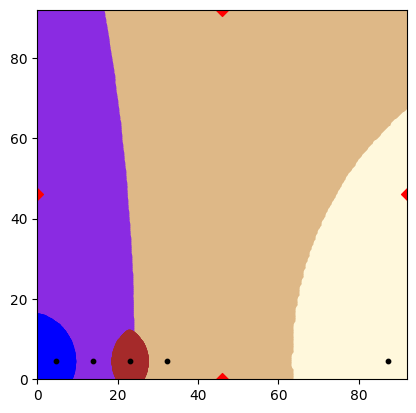}\\\hline
\end{tabular}
    \caption{Multiplicatively Weighted Voronoi diagrams of subjects 4 and 10.}
\label{Tabla: 3-sesiones}
\end{table}
\end{center}

\section{Conclusions}
Multiplicatively weighted Voronoi diagrams, applied in the context of SDBA, proved to be a valuable tool for visualizing and understanding various aspects of individual behavior. 

Each region in the weighted Voronoi diagram defines an RBR. The various organizations of these regions represent well-differentiated patterns of behavioral segmentation within the same experimental space. In other words, they depict distinctive spatial behavioral patterns related to the specific spatiotemporal features of the environment. Identifying these RBRs provides valuable information about behavior dynamics in relation to environmental conditions.

The weighted Voronoi diagrams evolve across experimental sessions, allowing the tracking of spatial patterns known as RBRs. The evolution of these patterns shows how behavior adapts to different conditions, offering crucial insights into a subject's behavioral adjustments over time and space.

The weighted Voronoi diagram highlights not only areas of close influence but also areas of distant influence. This aspect could help us identify not only preference patterns (i.e., RBR with approaching valence) but also those RBRs that tend to be avoided or have withdrawal valence. This feature could enrich the understanding of spatial variables and features related to a wide range of behavioral phenomena, incorporating not only preferences derived from the highest values of time spent in a given region but also aversive patterns indicated by the lowest values of time spent in a zone. It could provide a more comprehensive perspective on the dynamics of different behavioral phenomena.

The above observations highlight the utility of this tool for analyzing and visualizing spatial dynamics in behavioral studies. The ability to identify, track evolution, and distinguish spatial patterns in a perspicuous way, such as RBR, adds depth and precision to SDBA. Moreover, this capability could facilitate interpretation and informed decision-making in applied behavioral settings.

\end{document}